# A Novel Beam Tracking Approach for Preventing Beam Collapse

Xiaocun Zong[1], Fan Yang[1]*, Shenheng Xu[1], Maokun Li[1]
[1]*Department of Electronic Engineering, Tsinghua University*
Beijing 100084, China
E-mail: fan_yang@tsinghua.edu.cn

*Abstract*—To address the issue of beam collapse resulting from instantaneous instability during switch transitions in beam tracking, this paper proposes a novel beam switching method based on a row-by-row switching code table. The paper first establishes an abstract model of the beam tracking application scenario and introduces the reconfigurable intelligent surface (RIS) employed in this paper. Subsequently, simulations are conducted to compare the conventional direct beam switching method with the proposed row-by-row switching code table approach, thereby elucidating the advantages and limitations of the new method. In parallel, a RIS hardware platform is constructed in a microwave anechoic chamber for experimental validation. Both simulation and experimental results show that, by incorporating intermediate state transitions, the approach achieves beam tracking without beam collapse while incurring no significant gain loss. Finally, the paper discusses the applicability scope and potential scenarios for the proposed method. This research provides valuable insights for applications in mobile communications and radar detection.

## I. Introduction

In wireless communication systems, ensuring uninterrupted beam tracking during positional changes of a mobile terminal relative to a base station is critical for maintaining high-gain beam alignment and supporting high-speed communication. This issue is particularly significant in scenarios requiring robust and continuous connectivity. Existing literature has proposed various beam tracking schemes leveraging the Kalman filter algorithm [1], [2]; however, these approaches remain largely confined to simulation studies, lacking validation in practical applications.

In real-world engineering scenarios, beam tracking typically employs two primary strategies. When both the terminal and base station are equipped with positioning systems, beam pointing can be dynamically recalculated based on the terminal's real-time location, enabling precise beamforming and tracking. Alternatively, in cases where the terminal is distant or lacks a positioning system, beam tracking relies on variations in Received Signal Strength Indicator (RSSI) values. Significant RSSI fluctuations signal terminal movement, prompting a scan around the current beam position to identify the optimal RSSI and execute beam switching. While the aforementioned methods rely on direct beam switching, which can result in beam collapse.

In a RIS-assisted communication system, beam switching is achieved by altering the state of PIN diodes on the array elements [3]-[5]. However, the instantaneous transition of switch states introduces instability, leading to erratic beam pointing or beam collapse—a phenomenon where the beam temporarily disappears. Without a timeout mechanism in the communication protocol, beam collapse can disrupt communication process, necessitating beam re-formation and reconnection.

To address this problem, this paper investigates methods to ensure continuous high-gain reception during terminal movement along a known trajectory. Specifically, it proposes a solution based on a row-by-row switching code table to eliminate beam collapse caused by switching instability, thereby achieving seamless and continuous beam tracking. This research provides valuable insights for enhancing the reliability of RIS communication systems in mobile communications and related applications.

The paper is structured as follows: Section II introduces the fundamentals of RIS and models the beam tracking scenario. Section III proposes a novel row-by-row switching code table method, conducts simulation-based verification, and compares its performance with the conventional direct beam switching approach. Section IV describes the development of an RIS test system and presents experimental validation of the theoretical analysis, then discusses the applicable scenarios for the proposed method. Section V summarizes the key findings and contributions of the study.

## II. Modeling and RIS Parameters

### A. Scene Modeling

In a RIS-assisted communication system, where the RIS rotates on a turntable at a constant angular velocity of 10°/s while the terminal remains stationary until the RIS reaches a 10° rotation, this scenario is modeled as equivalent to the terminal moving relative to a fixed RIS. The terminal, initially positioned at (0°, 0°) on the RIS array, transitions to (10°, 0°) at the end of the rotation, located in the far field, 20 meters from the RIS base station, which can be seen in Fig. 1.

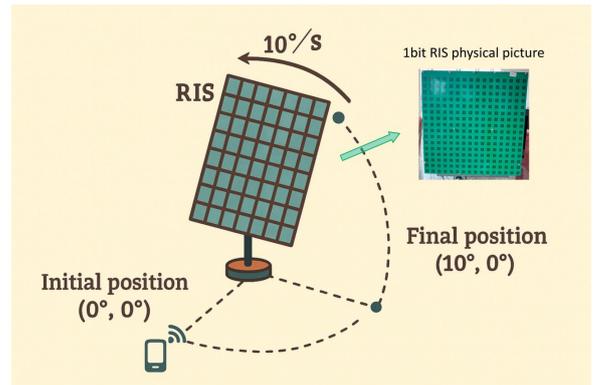

Fig. 1. Modeling of beam tracking.

## B. RIS Basic Parameters and Principles

The RIS employed in the experiment comprises a 1-bit 16 × 16 configuration, operating at a center frequency of 5.8 GHz, with each array element sized at $\lambda/2$. The amplitude and initial phase of each element are intrinsic characteristics determined by the RIS feed network, which is demonstrated in Fig. 2.

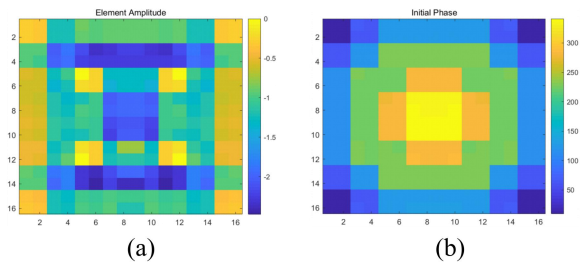

(a)  (b)

Fig. 2. RIS initial parameters: (a) the amplitude and (b) initial phase.

In RIS beamforming, the phase of each element is calculated to steer the beam toward a desired direction. Given the free space wave vector is $k$, the initial phase of the $mn$-th element is $\varphi_{mn\_init}$, and the beam direction is represented by the unit vector $\vec{u_0}$, the phase $\varphi_{mn}^{req}$ for each element can be expressed as:

$$\varphi_{mn}^{req} = \varphi_0 - k \cdot \vec{u_0} \cdot \vec{r_{mn}} - \varphi_{mn\_init} \quad (1)$$

The calculated phase $\varphi_{mn}^{req}$ for each element is quantized to $\varphi_{mn}$ using 1-bit quantization. According to the array principle, the radiation pattern of the RIS can be expressed as:

$$E(\theta, \varphi) = cos^{q_e} \sum_{m=0}^{M-1} \sum_{n=0}^{N-1} A_{mn} e^{j\varphi_{mn\_init}} e^{j\varphi_{mn}} e^{jk \cdot \vec{u_0} \cdot \vec{r_{mn}}} \quad (2)$$

The parameter $q_e$ is determined by the directivity of the individual array element, while $A_{mn}$ and $\varphi_{mn\_init}$ represent the amplitude and phase characteristics inherent to the RIS, as depicted in Fig. 2.

## III. A NOVEL SWITCHING CODE TABLE METHOD

### A. The Process of Row-by-Row Switching

Direct beam switching via code table transitions requires phase state changes across most array elements, leading to beam collapse during the switching process. To address this, the proposed method compares the code tables for the main beam 0° beam and the 10° beam, implementing beam switching by sequentially updating the code table row by row during the RIS rotation. As modifying the code table for a single row leaves other rows unaffected, the majority of array elements maintain their state, preserving the main beam's stability. This approach effectively prevents beam collapse, ensuring continuous high-gain beam tracking. The comparison of code table switching methods during beam switching is shown in Fig. 3.

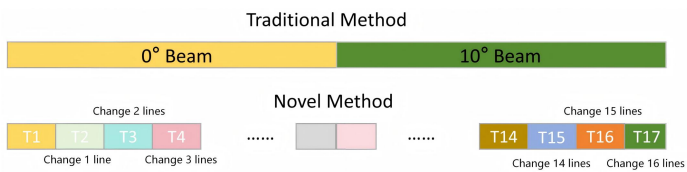

Fig. 3. Comparison of code table switching methods during beam switching

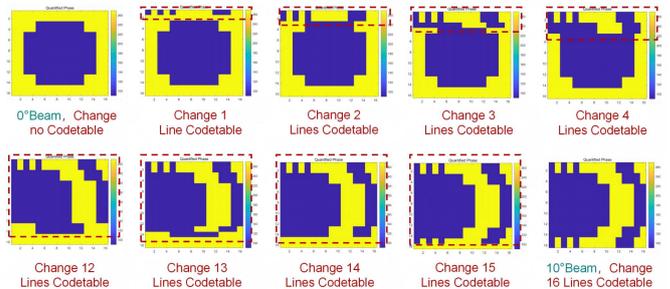

Fig. 4. Switch code table row by row from 0° to 10°.

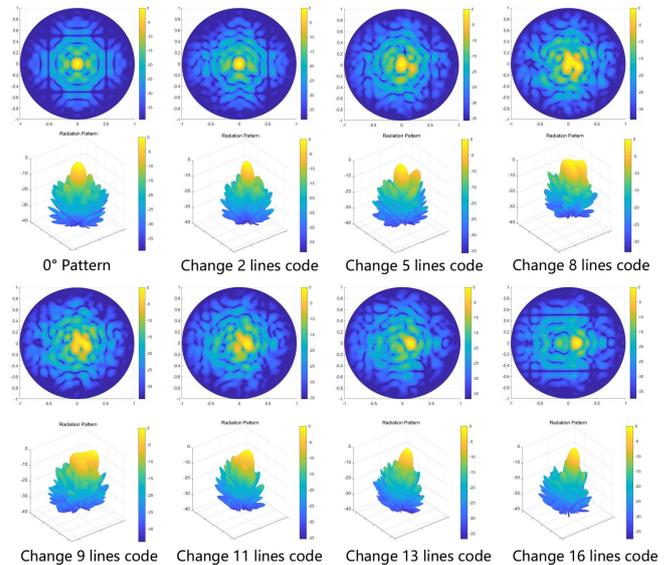

Fig. 5. The change process of the pattern when changing the code table row by row.

The study demonstrates this using a 16 × 16 RIS array as an example. As illustrated in Fig. 4, beam switching is accomplished by sequentially updating the code table row by row, ensuring a smooth beam transition. As the number of modified code table rows increases, the sidelobe level of the 0° beam becomes more pronounced. When approximately eight rows of the code table have been altered, the 16 × 16 array effectively splits into two 8 × 16 subarrays. However, due to the close alignment of their beams, they merge into a single wide beam. As additional rows are changed, the main lobe gradually shifts to 10°, with the sidelobe effect diminishing, ultimately yielding a stable 10° beam pattern, the change process of the pattern can be seen in Fig. 5. It can be seen that, the row-by-row code table switching method enables a gradual beam transition from 0° to 10°, avoiding abrupt jumps. The beam switching process is driven by row-by-row code table updates, synchronized with the turntable's constant rotation speed. During the 1-second rotation from 0° to 10°, the code table is updated 16 times, with one row changed approximately every 62 ms, achieving seamless beam switching.

### B. Simulation Comparison of Terminal Receiving Level

Beam tracking simulations were conducted based on the modeled scenario in Section II, evaluating the received signal level at the terminal. In the conventional beam switching method, beam switching at 500 ms induces beam collapse, disrupting signal stability. In contrast, the proposed row-by-row switching code table approach maintains a stable beam state, as most array elements remain unchanged during each switch. The simulation results of terminal received level during beam tracking can be seen in Fig. 6.

However, it should be noted that although the novel method allows the beam to slowly complete tracking and always be aligned with the terminal, row-by-row switching reduces the number of elements contributing to the main beam synthesis, leading to a slight reduction in main beam strength and the received signal level at the terminal. This trade-off must be carefully considered in practical applications.

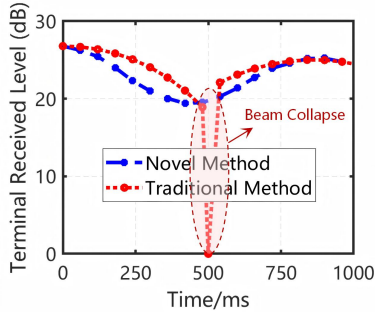

Fig. 6. Simulation results of terminal received level during beam tracking.

## IV. SYSTEM CONSTRUCTION AND TEST VERIFICATION

To evaluate the performance of beam tracking using the row-by-row code table switching method versus direct code table switching, experimental measurements were conducted in the B3 microwave anechoic chamber located in the Electronic Engineering Building at Tsinghua University, which is shown in Fig. 7. These results were subsequently compared with simulations to validate the proposed approach.

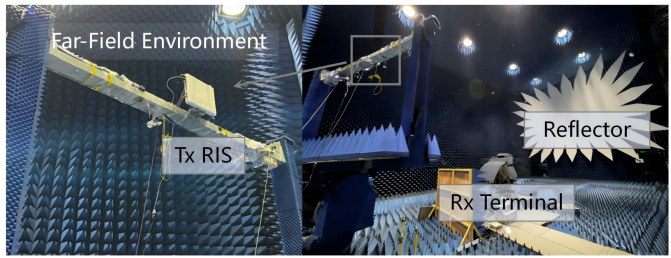

Fig. 7. Far-field environment test real shot.

Comparison of the received signal levels between experimental measurements and simulations in Fig. 8 reveals that the measured levels are approximately 4 dB lower than the simulated levels, though the trends of variation are nearly identical. This discrepancy is likely attributable to factors not accounted for in the simulation, such as insertion loss of individual elements (~1dB), environment effects, and antenna cover losses (~1dB). Comparative analyses of simulation and measurement data demonstrate that the row-by-row code table switching method enables high-gain beam tracking without beam collapse. While this approach results in a slight gain reduction compared to direct code table switching, the overall gain remains high. The consistency between measured and simulated results further validates the accuracy of the simulation methodology and its findings.

It is noteworthy that the proposed method is particularly effective for small-angle beam tracking. In such scenarios, the row-by-row code table switching approach ensures that the beams before and after the switch are closely aligned, merging into a single beam during the transition, thus avoiding the dual-beam phenomenon. This effectively mitigates beam collapse caused by state instability, enabling high-gain beam tracking. For large-angle beam switching, while the method still addresses beam collapse due to unstable states, the row-by-row switching process results in two distinct code table segments, each forming a separate beam, leading to a dual-beam phenomenon. Consequently, if the target terminal falls outside the beam width, the received signal level may decrease significantly. In conclusion, the row-by-row code table switching method is well-suited for small-angle beam tracking, provided the tracking is performed in a timely manner.

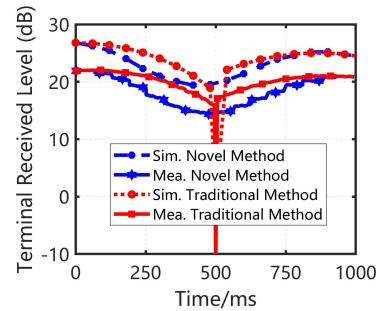

Fig. 8. Measurement and simulation results of terminal received level during beam tracking.

## V. CONCLUSIONS

During beam tracking, abrupt beam switching can lead to beam collapse, causing a significant drop in signal strength. To address this challenge, this paper proposes an innovative solution utilizing row-by-row code table switching. This method incrementally adjusts the code table configuration during beam switching, incorporating intermediate states to ensure a smooth transition process. By doing so, it effectively prevents beam collapse while maintaining high gain with minimal loss. The feasibility and effectiveness of this approach were validated through simulations and maesurements, demonstrating its applicability in real-world systems. This research provides valuable insights for applications in mobile communications and radar detection.

ACKNOWLEDGMENT

This work is supported by the National Key Research and Development Program of China under Grant No. 2023YFB3811501.